\documentclass{arxiv-preprint}

\usepackage{listings}
\usepackage{xcolor}
\usepackage{tikz}
\usepackage{amsmath}
\usepackage{amssymb}
\usepackage{amsfonts}
\usepackage{siunitx}
\usepackage{makecell}
\usepackage{array}
\usepackage{float}
\usepackage{tabularx}
\usepackage[T1]{fontenc}
\usepackage{booktabs}
\usepackage[backend=biber,style=authoryear-comp,sorting=nyt,sortcites=false,doi=false,url=true,natbib=true,
            maxcitenames=2,mincitenames=1,maxbibnames=10,minbibnames=1,giveninits=true,urldate=iso,seconds=true,
            uniquename=false,uniquelist=false,dashed=false]{biblatex}
\DefineBibliographyStrings{english}{%
  andothers = {\emph{et\addabbrvspace al}\adddot},
}
\AtEveryBibitem{%
  \clearfield{month}%
  \clearfield{issn}%
  \clearfield{pmid}%
  \clearfield{note}%
  \clearlist{publisher}%
  \clearlist{location}%
}

\newcounter{daggerfootnote}

\definecolor{codebg}{rgb}{0.97,0.97,0.97}
\definecolor{codegray}{rgb}{0.4,0.4,0.4}
\definecolor{codeblue}{rgb}{0.1,0.2,0.6}
\definecolor{codegreen}{rgb}{0.1,0.5,0.1}

\lstdefinestyle{mystyle}{
    backgroundcolor=\color{codebg},
    basicstyle=\ttfamily\small,
    keywordstyle=\color{codeblue}\bfseries,
    commentstyle=\color{codegreen},
    stringstyle=\color{orange!70!black},
    numberstyle=\ttfamily\small\color{codegray},
    numbers=left,
    numbersep=8pt,
    stepnumber=1,
    frame=single,
    rulecolor=\color{black!15},
    breaklines=true,
    breakatwhitespace=true,
    tabsize=4,
    showstringspaces=false,
    captionpos=b,
    keepspaces=true,
    columns=fullflexible
}

\begin{document}

\articletype{Paper}%

\title{A geometry-aware framework for electromagnetic fields in GATE~10}

\author{Marc~Ballestero~Ribó$^{1,2,*}$\orcid{0009-0008-2560-3293}, Carla~Winterhalter$^1$\orcid{0009-0008-4289-984X}, Antony~John~Lomax$^{2}$\orcid{0000-0002-0666-425X} and Sairos~Safai$^{1}$\orcid{0000-0002-0459-4534}}

\affil{$^1$Center for Proton Therapy, Paul Scherrer Institute, Villigen, Switzerland}

\affil{$^2$Department of Physics, ETH Zürich, Zürich, Switzerland}

\affil{$^*$Author to whom any correspondence should be addressed.}

\email{marc.ballestero-ribo@psi.ch}


\begin{abstract}
\noindent\textit{Objective.} An increasing number of medical-physics applications require the modelling of charged-particle transport in electromagnetic (EM) fields. We aim to provide GATE~10 users with a flexible mechanism for defining such fields in Monte Carlo simulations while preserving the native Geant4 particle-transport machinery.

\noindent\textit{Approach.} Our framework separates the handling of coordinates from EM field evaluation. Fields are attached to geometry volumes, defined in their local frame, and represented by analytic expressions, field maps, or arbitrary Python functions. They automatically follow the position and orientation of these volumes, including in nested and dynamic geometries. We validated the framework against native Geant4 transport for uniform, quadrupole, and sextupole fields, and tested coordinate transformations, nested volumes, and dynamic geometry. We also compared the computational cost of the three field representations.

\noindent\textit{Main results.} The framework reproduced the native Geant4 trajectories exactly for all tested field types with direct Geant4 counterparts. Coordinate transformations introduced only floating-point-level differences, while simulations with dynamically changing geometry reproduced the equivalent static configurations exactly. All three representations produced equivalent transport. Mapped fields added at most \SI{18}{\percent} to the single-threaded runtime of analytic fields, while Python functions required \num{6.9} times as much. The compiled representations scaled with the number of threads, whereas Python functions became slower as threads were added, reaching \num{455} times the analytic runtime at eight threads.

\noindent\textit{Significance.} The proposed framework adds flexible, geometry-aware EM field definitions to GATE~10 without modifying the underlying Geant4 transport machinery, and is part of the open-source GATE~10 codebase. Analytic and mapped fields provide efficient options for production simulations, while Python functions allow arbitrary spatial field dependencies for prototyping and exploratory applications. The common framework also provides a basis for extending GATE~10 with additional field representations in the future.
\end{abstract}

\section{Introduction}

Monte Carlo (MC) particle-transport simulations are an essential tool in research and treatment planning in radiotherapy, providing a high-fidelity description of radiation transport and energy deposition in complex, heterogeneous media. GATE, an open-source layer built on top of the Geant4 toolkit \citep{agostinelli2003,allison2006,allison2016}, has been widely adopted across imaging and dosimetry \citep{sarrut2014,sarrut2021,sarrut2022}. The latest version, GATE~10, replaces the legacy macro interface with an accessible Python-based interface while retaining the capabilities of Geant4 \citep{sarrut2026,krah2026}.

An increasing number of medical-physics applications require the modelling of charged-particle transport in electromagnetic (EM) fields. A prominent example is magnetic-resonance-guided radiotherapy (MRgRT), where magnetic fields deflect charged-particle trajectories and can therefore alter dose distributions: in photon therapy through the deflection of secondary electrons \citep{raaijmakers2005,raaijmakers2008}, and in proton therapy through the deflection of the primary beam itself \citep{schellhammer2017,schellhammer2018a}. Beyond MRgRT, dipoles and quadrupoles steer and focus beams in accelerators and treatment beamlines, where modelling beam losses requires charged-particle transport through these fields together with interactions in the surrounding material \citep{nevay2020}. Electric and magnetic fields also play a central role in radiation detectors, where they act directly on the secondary charged particles that determine the detector signal. For instance, in Faraday cups used for proton dosimetry, electric and magnetic fields are applied to control secondary-electron contributions, which can limit the accuracy of the collected charge unless corrected for \citep{winterhalter2021,ehwald2023}. These applications differ not only in the fields involved, but also in how fields relate to the simulation geometry: a single field can encompass an entire patient, a beamline might contain several independently positioned field-producing elements, and a detector may contain a field confined to a single component. A general and well-validated capability for EM field transport in a medical-physics Monte Carlo framework is therefore of broad interest.

Geant4 provides a general framework for defining EM fields and for tracking charged particles through them. GATE has historically provided access to a subset of these capabilities through interfaces shaped by the applications that motivated them. Uniform magnetic fields were introduced in early versions of GATE for emission-tomography studies in which a single field encompasses the whole scanner, and were subsequently used to quantify the effect of magnetic fields on positron range in PET/MR \citep{soultanidis2011,barati2021}. Spatially varying fields were later introduced in GATE~8 through three-dimensional field maps supplied as look-up tables and interpolated over the extent of the grid \citep{padilla-cabal2020}. These capabilities were retained in GATE~9 and used, for example, to commission a beamline with an integrated dipole for MR-guided proton and carbon-ion radiotherapy \citep{fuchs2023}. However, neither interface was ported to GATE~10 during the transition to the Python-based architecture. This created a need to reintroduce EM field transport in GATE~10, and an opportunity to provide a more general mechanism for defining fields in relation to the simulation geometry, while preserving the native Geant4 particle-transport machinery.

In this work, we introduce a framework for defining EM fields in GATE~10 and associating them directly with simulation volumes. The framework separates the handling of coordinates from EM field evaluation, allowing the same geometry-dependent coordinate treatment to be used with different field representations. Three interchangeable approaches are provided: analytic fields for closed-form descriptions, mapped fields for fields defined on regular grids, and Python functions for arbitrary user-defined spatial dependencies. Fields are defined in the local coordinate system of their associated volume and automatically follow its position and orientation, including when the geometry is nested or dynamically modified. We validate the framework against native Geant4 transport, test its behaviour under translations, rotations, and nested and dynamic geometries, and compare the computational cost of the three field representations.

\section{Methods}
\label{sec:methods}

Our framework integrates user-defined EM fields into GATE~10 while leaving particle transport entirely to the native Geant4 machinery. It separates the field-evaluation process into two independent components: coordinate transformation and field evaluation. Sections~\ref{sec:transport} to \ref{sec:userview} describe the formulation and implementation of this approach. Section~\ref{sec:validation} describes the tests used to validate the framework.

\subsection{Charged-particle tracking in Geant4}
\label{sec:transport}

A track in a Monte Carlo simulation advances in discrete \emph{transport steps}. The transport step length is set by the particle--medium interaction properties and truncated at geometric boundaries, so that energy depositions are assigned to the correct volume and cross-sections are updated when an interface is crossed.

When a charged particle travels through a region of space containing an EM field, its trajectory is bent by the Lorentz force, according to the equation of motion
\begin{equation} \label{eq:eqmotion}
    \frac{d\mathbf{r}}{ds} = \frac{\mathbf{p}}{|\mathbf{p}|}, \qquad
    \frac{d\mathbf{p}}{ds} = \frac{q}{|\mathbf{p}|}\left(\, \mathbf{p}\times\mathbf{B}
    + \frac{|\mathbf{p}|}{v}\,\mathbf{E} \,\right),
\end{equation}
where $\mathbf{r}$ is the position vector, $\mathbf{p}$ the momentum vector, $s$ the path length, $q$ the charge, $v$ the speed, $\mathbf{B}$ the magnetic field, and $\mathbf{E}$ the electric field. Accordingly, Geant4 replaces the straight-line transport step by a numerical integration of \eqref{eq:eqmotion}, generally using an adaptive Runge--Kutta integration method. This subdivides the transport step into \emph{integration substeps}, whose length is adapted to keep the integration error within a prescribed tolerance. Each substep evaluates the right-hand side of \eqref{eq:eqmotion} at several intermediate points, one per stage of the integration method. Geant4 then approximates the resulting curved path by a series of straight segments, or ``chords'', which is what allows the geometry navigator to find boundary intersections. A candidate intersection is finally refined to within a user-defined tolerance.

Geant4 performs this transport through a chain of cooperating classes \citep[Sec.~4.3]{geant4manual}. An \emph{equation of motion} supplies the right-hand side of \eqref{eq:eqmotion}. An \emph{integration stepper} advances the state by one integration substep and estimates the local error, and an \emph{integration driver} adapts the substep length to keep that error within tolerance. A \emph{chord finder} locates boundary intersections using the chord approximation. Finally, a \emph{field manager} binds all the classes above together with the \emph{field object}, which holds the field definition. The field object exposes this definition through a single method, \texttt{GetFieldValue()}, which receives a query point in world coordinates and returns the field vector at that point, also in world coordinates. This method is called once per integration stage, so a single transport step may trigger several field queries.

The field manager is attached to a \emph{logical volume}, the reusable template describing the shape of a solid and its material, which may be instantiated as one or more \emph{physical volumes}, each being a placement of that template in the simulation geometry, thus carrying its own local coordinate frame. A single field object therefore serves every placement of its host logical volume at once.

Two families of numerical parameters govern the tracking accuracy. Table~\ref{tab:fieldparams} lists them, together with their default values as implemented in GATE~10, which are inherited from Geant4~11.4.2. The first family controls the numerical integration of \eqref{eq:eqmotion}, and hence the stepper and the integration driver; the second controls the intersection of the curved trajectory with volume boundaries, and hence the chord finder. All of them are exposed to the user, so they can be tuned to the needs of a particular simulation.

\begin{table}[htbp]
\centering
\caption{User-exposed parameters controlling the integration of the equation of motion and the intersection of the curved trajectory with volume boundaries, with their default values in GATE~10.}
\label{tab:fieldparams}
{\small
\begin{tabularx}{\linewidth}{@{}>{\raggedright\arraybackslash\leftskip=1em}X l l@{}}
\toprule
\multicolumn{1}{@{}l}{Quantity controlled} & GATE~10 keyword & Default value \\
\midrule
\multicolumn{3}{@{}l}{\textit{Integration of the equation of motion}} \\
    Numerical method used to integrate \eqref{eq:eqmotion}
      & \texttt{stepper}$^{a}$ & \texttt{DormandPrince745} \\
    Integration substep length below which the accuracy requirement is no longer enforced
      & \texttt{step\_minimum} & \SI{e-2}{\milli\metre} \\
    Position error accepted over one transport step, from which the relative tolerance $\varepsilon$ is derived
      & \texttt{delta\_one\_step} & \SI{e-2}{\milli\metre} \\
    Lower bound on $\varepsilon$
      & \texttt{min\_epsilon\_step} & \num{5e-5} \\
    Upper bound on $\varepsilon$
      & \texttt{max\_epsilon\_step} & \num{e-3} \\
\addlinespace
\multicolumn{3}{@{}l}{\textit{Intersection with volume boundaries}} \\
    Maximum sagitta between a chord and the true curved path
      & \texttt{delta\_chord} & \SI{0.25}{\milli\metre} \\
    Accuracy of the computed boundary-crossing point
      & \texttt{delta\_intersection} & \SI{e-3}{\milli\metre} \\
\bottomrule
\end{tabularx}
}
\par\smallskip
{\footnotesize\raggedright
$^{a}$Admissible values are \texttt{DormandPrince745} (default), \texttt{ClassicalRK4}, \texttt{CashKarpRKF45}, \texttt{BogackiShampine45}, \texttt{BogackiShampine23}, \texttt{DormandPrinceRK56}, and \texttt{DormandPrinceRK78}. For purely magnetic fields \texttt{NystromRK4} and \texttt{ExactHelixStepper} are also available. See \citep[Sec.~4.3.3]{geant4manual} for guidance on stepper selection.\par}
\end{table}

The tolerance required from the integration is expressed as a relative bound $\varepsilon$, computed once per transport step as the ratio of \texttt{delta\_one\_step} to the proposed transport step length, and then clamped to [\texttt{min\_epsilon\_step}, \texttt{max\_epsilon\_step}]. Within that transport step, the integration driver estimates the local error of each substep, compares it with $\varepsilon$, and subdivides the substep until the estimate is below tolerance. To avoid unbounded subdivisions, the accuracy requirement is relaxed for substeps shorter than \texttt{step\_minimum}.

The sagitta of each chord, i.e. the maximum distance between the chord and the arc it replaces, is bounded by \texttt{delta\_chord}. Once a candidate crossing has been found, \texttt{delta\_intersection} sets how precisely the crossing point is calculated.

Of the machinery described above, only the field object depends on how a field is defined; every other component is generic. Therefore, \texttt{GetFieldValue()} is the single interface through which a field definition can enter Geant4 transport.

\subsection{Implementation design}
\label{sec:design}

Our framework intervenes at a single point of the chain described above: it overrides \texttt{GetFieldValue()} in the field object and leaves the rest of the tracking machinery unchanged. The equation of motion, integration stepper, integration driver, and chord finder are therefore the native Geant4 ones, and any field defined through our framework is transported exactly as a native Geant4 field would be.

Three design goals shaped what happens inside the override. First, the user should define a field in the natural local frame of the volume that hosts it, while Geant4 queries it in world coordinates. Second, common fields with a closed form and fields supplied as field maps should be evaluated entirely in compiled code, since the field is queried at every stage of every integration substep, which makes field evaluation one of the most frequently executed operations in a simulation. Third, when neither of these cases applies, the user should still be able to supply the field as an ordinary Python function that encodes an arbitrary spatial dependence.

The override therefore separates coordinate handling from field evaluation. Coordinate handling is the same for every field, whereas field evaluation depends on how the field was defined, so we delegate it to an interchangeable component, which we call a \emph{field provider}.

\begin{figure}[tbp]
 \centering

\usetikzlibrary{arrows.meta,backgrounds,calc,fit,positioning}

\definecolor{g4line}{HTML}{4F4F4F}
\definecolor{g4band}{HTML}{F4F4F4}
\definecolor{g4fill}{HTML}{E6E6E6}
\definecolor{gbline}{HTML}{1F5C8B}
\definecolor{gbband}{HTML}{EDF3F9}
\definecolor{gbfill}{HTML}{D5E3EF}
\definecolor{pyline}{HTML}{B15A00}
\definecolor{pyband}{HTML}{FCF3E8}
\definecolor{keyline}{HTML}{aa0080}
\definecolor{hotred}{HTML}{A81F14}

\providecommand{\stepnum}[1]{%
  \tikz[baseline={([yshift=-0.55ex]n.center)}]{%
    \node[circle, draw=., fill=white, line width=0.5pt, inner sep=0.3pt,
          minimum size=2.5mm, font=\scriptsize\bfseries] (n) {#1};}%
}

\begin{tikzpicture}[
  x=1mm, y=1mm, font=\footnotesize,
  >={Stealth[length=1.7mm,width=1.4mm]},
  g4box/.style   ={draw=g4line, fill=white, rounded corners=1pt, align=center,
                   inner xsep=3pt, inner ysep=2.6pt},
  pybox/.style   ={draw=pyline, fill=white, rounded corners=1.5pt,
                   align=center, inner xsep=3pt, inner ysep=2.8pt},
  inner/.style   ={draw=gbline, line width=0.9pt, fill=gbfill,
                   rounded corners=1.5pt, align=center, inner ysep=2.8pt},
  arr/.style     ={->, shorten <=0pt, shorten >=0pt},%
  own/.style     ={arr, draw=g4line},%
  rt/.style      ={arr, line width=1.1pt,%
                   -{Stealth[length=2.1mm,width=1.7mm]}},%
  key/.style     ={rt, draw=keyline},%
  bld/.style     ={arr, draw=pyline, line width=0.8pt},%
  flow/.style    ={rt, draw=gbline},%
  tag/.style     ={font=\scriptsize\bfseries\sffamily, inner sep=0pt,
                   rotate=90, anchor=center},
  lbl/.style     ={font=\scriptsize, align=center, inner sep=1.5pt},
  tny/.style     ={font=\tiny, align=center, inner sep=1pt},
  dim/.style     ={font=\scriptsize, text=black!58, align=center, inner sep=1.5pt},
  cls/.style     ={font=\tiny\ttfamily, align=center, inner sep=0pt},
  grp/.style     ={draw=g4line!70, densely dashed, rounded corners=2pt},
]

\begin{scope}[on background layer]
  \fill[g4band] (0,-2)    rectangle (153,-68);
  \fill[gbband] (0,-68)   rectangle (153,-113);
  \fill[pyband] (0,-113)  rectangle (153,-152);
\end{scope}
\draw[white, line width=0.7pt] (0,-113) -- (153,-113);

\node[tag, text=g4line] at (5,-35)   {Geant4};
\node[tag, text=gbline] at (5,-90.5) {GATE \textendash\ C\texttt{++}};
\node[tag, text=pyline] at (5,-132)  {GATE \textendash\ Python};

\begin{scope}[xshift=3mm]
\draw[g4line] (-3,-68) -- (46,-68);
\draw[g4line] (118,-68) -- (150,-68);

\node[g4box, text width=32mm] (lv) at (27.5,-8) {logical volume};
\node[g4box, text width=32mm] (fm) at (91,-8)   {field manager};
\draw[own] (lv) -- (fm);

\draw[own] (lv) -- (27.5,-19);
\draw[g4line, fill=white, rounded corners=1pt] (11,-19)   rectangle (44,-25.5);
\draw[g4line, fill=white, rounded corners=1pt] (13.5,-21.5) rectangle (46.5,-28);
\draw[g4line, fill=white, rounded corners=1pt] (16,-24)   rectangle (49,-30.5);
\node at (32.5,-27.25) {physical volume};

\node[g4box, text width=32mm] (nav) at (27.5,-50) {navigation history};
\draw[own,dashed] (27.5,-30.5) -- (nav);

\draw[rt, draw=g4line] (nav.south) -- (27.5,-71) -- (46,-71);
\node[dim, font=\scriptsize\bfseries, text width=22mm, align=right, anchor=east] at (25,-61)
  {\stepnum{2}\ read the world\,$\to$\,local\\[0.6pt] transform, per query};

\draw[grp] (71,-14) rectangle (111,-46);
\node[dim] at (91,-17.5) {\itshape tracking and integration \\ \itshape infrastructure};
\node[g4box, text width=34mm] (cf) at (91,-23.5) {chord finder};
\node[g4box, text width=34mm] (dr) at (91,-30)   {integration driver};
\node[g4box, text width=34mm] (st) at (91,-36.5) {stepper};
\node[g4box, text width=34mm] (eq) at (91,-43)   {equation of motion};
\draw[own] (fm) -- (91,-14);
\draw[own] (cf) -- (dr);
\draw[own] (dr) -- (st);
\draw[own] (st) -- (eq);

\draw[key] (89,-46) -- (89,-58);
\node[lbl, font=\scriptsize\bfseries, align=right, anchor=east, text=keyline] at (87,-52)
  {\stepnum{1}\ query at $\mathbf{p}_\mathrm{w}$};
\draw[key] (93,-58) -- (93,-46);
\node[lbl, font=\scriptsize\bfseries, anchor=west, text=keyline] at (95,-52)
  {\stepnum{6}\ return $\mathbf{F}_\mathrm{w}(\mathbf{p}_\mathrm{w})$};

\begin{scope}
  \clip[rounded corners=2pt] (46,-58) rectangle (118,-78);
  \fill[white]  (46,-58) rectangle (118,-78);
  \fill[g4fill] (46,-58) rectangle (118,-68);
\end{scope}
\draw[g4line, densely dashed] (46,-68) -- (118,-68);
\draw[rounded corners=2pt, line width=0.9pt, draw=g4line]
  (46,-58) rectangle (118,-78);

\node[align=center, text width=68mm] at (82,-63)
  {\textbf{Geant4 field object}\\[-1pt]
   {\scriptsize\color{black!58}\itshape the Geant4 class that declares
    \textcolor{keyline}{\upshape\ttfamily GetFieldValue()}}};
\node[align=center, text width=68mm] at (82,-73)
  {\textbf{GATE field object}\\[-1pt]
   {\scriptsize\color{black!58}\itshape derives from it and overrides
    \textcolor{keyline}{\upshape\ttfamily GetFieldValue()}}};

\node[inner, text width=24mm, anchor=north] (fld) at (85,-87)
  {$\mathbf{F}_\ell = f(\mathbf{p}_\ell)$};
\node[font=\scriptsize\bfseries, text=gbline, anchor=south] at (85,-86.7)
  {\stepnum{4}\ evaluate field in the local frame};

\draw[flow] (46,-74) -- (40,-74) |- (fld.west);
\node[font=\scriptsize\bfseries, align=right, text width=23mm, text=gbline,
      anchor=east] at (38,-83)
  {\stepnum{3}\ apply the transform,\\[0.6pt]
   $\mathbf{p}_\mathrm{w} \mapsto \mathbf{p}_\ell$};

\draw[flow] (fld.east) -| (124,-74) -- (118,-74);
\node[font=\scriptsize\bfseries, align=left, text=gbline, anchor=west] at (126,-82)
  {\stepnum{5}\ transform\\[0.6pt] $\mathbf{F}_\ell \mapsto \mathbf{F}_\mathrm{w}$};

\node[g4box, draw=gbline, text width=30mm, anchor=north] (cxx) at (58,-103)
  {\textbf{compiled \texttt{C++}}\\[-1pt]
   {\scriptsize\color{black!58}\itshape built once at setup}};
\draw[flow] (58,-103) -- (58,-99) -- (81.5,-99) -- (81.5,-92.2);

\draw[rt, draw=pyline] (114,-121) -- (114,-99) -- (88.5,-99) -- (88.5,-92.2);

\fill[white] (114,-113) circle (2.9mm);
\draw[hotred, line width=1.0pt] (114,-113) circle (2.75mm);
\draw[hotred, line width=0.8pt] (113.1,-112.45)
  arc[start angle=180, end angle=0, radius=0.9mm];
\fill[hotred, rounded corners=0.25mm] (112.6,-114.45) rectangle (115.4,-112.45);
\fill[white] (114,-113.25) circle (0.28mm);
\fill[white] (113.89,-113.25) -- (113.79,-114.05) -- (114.21,-114.05)
          -- (114.11,-113.25) -- cycle;
\node[font=\scriptsize\bfseries, text=hotred, align=left, anchor=west] at (118,-107.5)
  {Global interpreter\\[-1pt] lock acquired\\[-1pt] on every call};

\draw[pyline, line width=0.8pt, densely dashed] (34,-121) -- (34,-115) -- (79,-115) -- (79,-121);
\draw[bld, densely dashed] (58,-115) -- (cxx.south);

\draw[densely dashed, rounded corners=2pt, draw=pyline!70]
  (11.5,-118) rectangle (142,-148);
\node[lbl, fill=pyband, text=pyline, inner xsep=3pt, anchor=west] at (12.5,-118)
  {\itshape field providers};

\node[pybox, text width=30mm, anchor=north] at (34,-121)  {\textbf{analytic}};
\node[pybox, text width=30mm, anchor=north] at (79,-121)  {\textbf{mapped}};
\node[pybox, text width=30mm, anchor=north] at (124,-121) {\textbf{callback}};

\node[anchor=north west, font=\fontsize{6.4}{8}\selectfont\ttfamily, align=left] at (16.5,-126.5)
  {{\tiny\rmfamily\textbullet}\ UniformMagneticField\\
   {\tiny\rmfamily\textbullet}\ UniformElectricField\\
   {\tiny\rmfamily\textbullet}\ UniformElectroMagneticField\\
   {\tiny\rmfamily\textbullet}\ QuadrupoleMagneticField\\
   {\tiny\rmfamily\textbullet}\ SextupoleMagneticField};
\node[anchor=north west, font=\fontsize{6.4}{8}\selectfont\ttfamily, align=left] at (61.5,-126.5)
  {{\tiny\rmfamily\textbullet}\ MappedMagneticField\\
   {\tiny\rmfamily\textbullet}\ MappedElectricField\\
   {\tiny\rmfamily\textbullet}\ MappedElectroMagneticField};
\node[anchor=north west, font=\fontsize{6.4}{8}\selectfont\ttfamily, align=left] at (106.5,-126.5)
  {{\tiny\rmfamily\textbullet}\ CustomMagneticField\\
   {\tiny\rmfamily\textbullet}\ CustomElectricField\\
   {\tiny\rmfamily\textbullet}\ CustomElectroMagneticField};

\end{scope}

\end{tikzpicture}
 \caption{The framework is a single override of one Geant4 method, \texttt{GetFieldValue()}. A field is registered by attaching a field manager to a logical volume (\emph{top}), so one field object serves every physical placement of that volume. The field manager owns the chord finder, integration driver, stepper and equation of motion. Field queries are emitted once per integration stage of each integration substep. That is the only point at which the field definition enters the tracking machinery. The GATE field object (\emph{middle}) derives from the corresponding Geant4 field class and overrides \texttt{GetFieldValue()}, wrapping the evaluation in the shared coordinate pipeline. The evaluation of the field in local coordinates is supplied by one of three interchangeable \emph{field providers}, exposed to the user as eleven concrete types (\emph{bottom}).}
 \label{fig:architecture}
\end{figure}
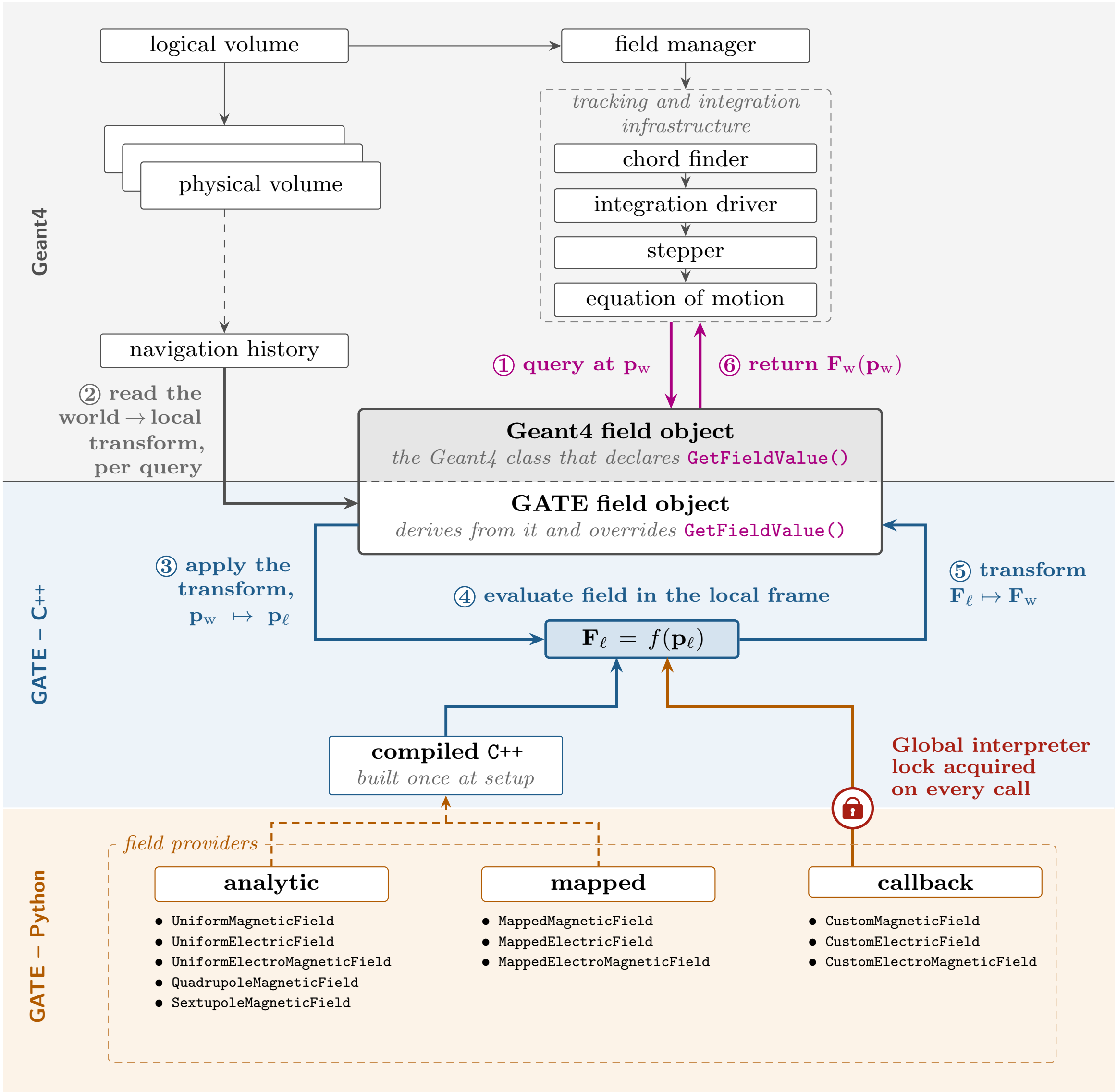

Figure~\ref{fig:architecture} illustrates the resulting design. Section~\ref{sec:pipeline} describes the change of coordinates, Section~\ref{sec:providers} covers the three field providers, and Section~\ref{sec:userview} then presents the user-facing Python interface.

\subsection{The coordinate pipeline}
\label{sec:pipeline}

The first part of the override transforms the query from the world frame used by Geant4 to the local frame of the volume to which the field is attached. We denote this query point in the world frame by $\mathbf{p}_\mathrm{w}$, and the corresponding position in the local frame of the host volume by $\mathbf{p}_\ell$. Once the host placement has been identified, the point $\mathbf{p}_\mathrm{w}$ is transformed to its local frame to obtain $\mathbf{p}_\ell$. The field is then evaluated by the corresponding provider as $\mathbf{F}_\ell = f(\mathbf{p}_\ell)$, where $\mathbf{F}_\ell$ is the field vector in the local frame. The result is transformed back to the world frame, giving $\mathbf{F}_\mathrm{w}$, which is returned to Geant4. Because this coordinate handling is independent of the field definition, it is implemented once in a \texttt{C++} class template shared by all field providers.

The host placement is resolved from Geant4's navigation history. A physical volume is placed inside a \emph{mother} volume, of which it is a \emph{daughter}, at its own position and orientation in the mother's frame, so the geometry is a tree rooted at the \emph{world}, the volume containing all others. The transform from world coordinates to the local frame of any physical volume is the composition of the transforms along the chain of ancestors linking it to the world. Geant4 maintains exactly that chain as a track advances through the geometry: the navigator records the physical volumes from the world down to the volume currently occupied by the track. This chain is the \emph{navigation history} of the track, and each of its levels stores the corresponding physical volume and its world-to-local transform. At each field query, the chain is searched from the deepest level towards the world until the physical volume whose logical volume matches the field's host logical volume is found. Its stored transform is used to transform $\mathbf{p}_\mathrm{w}$ to $\mathbf{p}_\ell$. After evaluating the provider, the inverse rotation is applied to $\mathbf{F}_\ell$ to obtain $\mathbf{F}_\mathrm{w}$.

A field defined for a host logical volume is inherited by its daughter volumes unless a daughter defines a field of its own. Thus, a track inside a daughter without its own field is governed by the field of the nearest ancestor to which a field is attached. Since the field is evaluated in the host volume's local frame, its orientation remains defined relative to that volume rather than the daughter containing the track. A rotation of the host volume therefore rotates the field together with the volume, while preserving its orientation relative to the host throughout its daughters.

This mechanism also supports dynamic geometry in GATE~10, where a simulation is executed as a sequence of runs between which dynamic actors may translate or rotate volumes \citep{sarrut2026}. Because the navigation history is rebuilt from the active geometry at the beginning of each run, the transform retrieved during field evaluation automatically reflects the placement of the host volume for the current run.

\subsection{Field providers}
\label{sec:providers}

The evaluation of the field in local coordinates is supplied by a \emph{field provider} (Figure~\ref{fig:architecture}, bottom band). Three providers are available, corresponding to the three ways in which a user may define a field: as a closed-form analytic expression, as a tabulated field map, or as an arbitrary Python function. All three expose the same evaluation interface to the coordinate pipeline, while differing in how they represent and evaluate the field internally. The data the user supplies to configure them is covered in Section~\ref{sec:userview}.

\paragraph{Analytic fields.} The analytic provider represents fields for which a common closed form is available. We implement uniform magnetic and electric fields, ideal magnetic quadrupoles and sextupoles, and a uniform EM field. The first four are taken from the Geant4 catalogue of pre-defined field objects, while the last is provided by GATE. At runtime, the provider delegates field evaluation to the corresponding compiled field object. This provider therefore introduces little overhead, making analytic fields the fastest option, at the cost of being restricted to the available closed-form implementations.

\paragraph{Mapped fields.} The mapped provider represents fields that are known numerically, such as the output of a finite-element solver or a measured field map. A mapped field is defined by a regular Cartesian grid of nodes, each carrying a field vector, tabulated in the local frame of the host volume. At setup, the Python layer derives the grid origin and spacing from the node positions and transfers the data to the \texttt{C++} layer, where each field component is stored as a contiguous array in lexicographic node order. A query can therefore be evaluated through an index computation followed by interpolation of the neighbouring nodes.

Two interpolation schemes are available. \emph{Nearest-neighbour} interpolation returns the field of the closest node. \emph{Trilinear} interpolation returns the weighted average of the eight nodes of the enclosing cell,
\begin{equation}\label{eq:trilinear}
    \mathbf{F}(\mathbf{p}) = \sum_{c=0}^{7} w_c(\mathbf{p})\,\mathbf{F}_c,
    \qquad \sum_c w_c = 1,
\end{equation}
with each weight $w_c$ being the product of the fractional coordinates, or their complements, within the cell. Outside the grid, the query point is clamped to the nearest face, edge or vertex, so that the field is held constant beyond the map rather than extrapolated.

\paragraph{Callback fields.} The callback provider allows the field to be defined by an arbitrary Python function of the local coordinates of the query point. The underlying field object is created using the \texttt{pybind11} \emph{trampoline} pattern that GATE~10 uses for callbacks from Geant4 \citep{krah2026}. Any field expressible in Python is therefore admissible, which makes this a particularly useful provider for prototyping and exploratory work. This flexibility comes at a substantial runtime cost: each query crosses the \texttt{C++}/Python boundary and acquires the global interpreter lock, serialising field evaluation across worker threads even in a multi-threaded simulation. We quantify this cost in Section~\ref{sec:results}.

The provider choice also affects the ability to serialise a simulation for archiving and replay. GATE~10 serialises simulations to JSON \citep{sarrut2026}, allowing the simulation configuration to be stored and reconstructed. Analytic and mapped fields can be represented in this format, whereas callback fields cannot, because arbitrary Python functions do not have a canonical JSON representation. Consequently, simulations containing callback fields cannot currently be fully serialised and replayed from the JSON configuration alone, and an exception is raised when attempting to do so.

\subsection{The user-facing Python interface}
\label{sec:userview}

At the Python level, a field is an ordinary simulation object, inheriting from the GATE~10 \texttt{GateObject} base class \citep{krah2026}. The user creates a field with a name, configures it through attributes, and attaches it to a volume with \texttt{volume.add\_field(...)}. A logical volume can have at most one attached field, reflecting the fact that Geant4 binds a single field manager to a logical volume. Electric and magnetic fields that coexist in the same region are therefore described by one joint EM field object rather than by two separate fields.

A field attached to a logical volume automatically applies to all of its physical placements, with the field evaluated in the local frame of each placement through the coordinate pipeline of Section~\ref{sec:pipeline}. Conversely, the same Python field object can be attached to multiple distinct volumes, allowing one field definition to be reused throughout the geometry. At setup, GATE creates the corresponding internal field object for each attachment. Integration and boundary tolerances, i.e. the parameters of Table~\ref{tab:fieldparams}, are configured as field attributes along with the field data.

The three providers of Section~\ref{sec:providers} are exposed to the user as eleven concrete classes, listed in Table~\ref{tab:gate_fields}. The table summarises the available field types and the field data required to configure each of them. Examples for all possible configurations are provided in the GATE user guide \citep{gate10manual}.

\begin{table}[ht]
\centering
\footnotesize
\caption{The eleven field types available through the GATE~10 Python interface, and the data the user supplies for each. All field data is expressed in the local frame of the host volume.}
\label{tab:gate_fields}
\begin{tabularx}{\textwidth}{
    @{}>{\hspace{1em}}l
    >{\raggedright\arraybackslash}X@{}
}
\toprule
\multicolumn{1}{@{}l}{\textbf{Python class}} & \textbf{Field data supplied by the user} \\
\midrule
\multicolumn{2}{@{}l}{\textit{Analytic provider}} \\
\texttt{UniformMagneticField} & Constant field vector \\
\texttt{UniformElectricField} & Constant field vector \\
\texttt{UniformElectroMagneticField} & Constant electric and magnetic field vectors \\
\texttt{QuadrupoleMagneticField} & Field gradient \\
\texttt{SextupoleMagneticField} & Field gradient \\
\addlinespace
\multicolumn{2}{@{}l}{\textit{Mapped provider}} \\
\texttt{MappedMagneticField} & Field sampled on a Cartesian grid, plus interpolation scheme \\
\texttt{MappedElectricField} & Field sampled on a Cartesian grid, plus interpolation scheme \\
\texttt{MappedElectroMagneticField} & Electric and magnetic fields sampled on independent Cartesian grids, plus interpolation scheme \\
\addlinespace
\multicolumn{2}{@{}l}{\textit{Callback provider}} \\
\texttt{CustomMagneticField} & Python function of position returning $(B_x,B_y,B_z)$ \\
\texttt{CustomElectricField} & Python function of position returning $(E_x,E_y,E_z)$ \\
\texttt{CustomElectroMagneticField} & Python function of position returning $(B_x,B_y,B_z,E_x,E_y,E_z)$ \\
\bottomrule
\end{tabularx}
\end{table}

For analytic fields, the configuration consists of the parameters of the corresponding closed-form expression. Uniform fields are specified by their field vector, while quadrupoles and sextupoles by their field gradient.

Listing~\ref{lst:minimal_example} shows a minimal example: a uniform magnetic field defined in a box. The three steps it follows, creating the field, setting its data and tolerances, and attaching it to a volume, are the same for every class of Table~\ref{tab:gate_fields}; only the field data differs.

\begin{lstlisting}[language=Python, caption={Minimal example: a uniform magnetic field defined in the local frame of a volume.}, label={lst:minimal_example}]
import opengate as gate
from opengate.geometry import fields

sim = gate.Simulation()
tesla = gate.g4_units.tesla
cm = gate.g4_units.cm

box = sim.add_volume("Box", "magnet")
box.size = [50 * cm, 50 * cm, 50 * cm]
box.material = "G4_Galactic"

field = fields.UniformMagneticField(name="B")
field.field_vector = [0, 1 * tesla, 0]  # in the local frame of `box`
field.stepper = "ClassicalRK4"
field.delta_chord = 1e-3 * cm

box.add_field(field)
\end{lstlisting}

Mapped fields are configured by supplying the field data through the \texttt{field\_matrix} attribute as an $N\times 6$ matrix whose columns are the three node coordinates and the three field components, all expressed in the local frame of the host volume. The interpolation scheme is selected through the \texttt{interpolation} attribute. For a mapped EM field, the electric and magnetic components are supplied separately through the \texttt{field\_matrix\_E} and \texttt{field\_matrix\_B} attributes, whose grids can be independent of each other. The supplied matrices are validated at setup: the spacing must be uniform along each axis, every combination of sampled coordinates must be present, and each axis must carry at least two distinct values. A matrix that violates any of these constraints raises an exception before the simulation starts. Tabulated field data is normally produced by a field solver or by a measurement, so it is loaded from a file rather than written inline, and the field is then configured and attached exactly as in Listing~\ref{lst:minimal_example}.

Callback fields are configured by assigning a Python function to the \texttt{field\_function} attribute. The function receives the local coordinates of the query point, and returns the field components in the local frame. Purely electric or magnetic fields return three components, while a combined EM field returns six, the first three being the magnetic components and the last three the electric ones. Listing~\ref{lst:custom_example} illustrates this signature and return convention with the field of an infinite straight wire.

\begin{lstlisting}[language=Python, caption={A custom field defined as a Python function of the local coordinates: the field of an infinite straight wire along the local $z$ axis, normalised to \texttt{B\_a} at a radius \texttt{a}.}, label={lst:custom_example}]
mm = gate.g4_units.mm

a = 10 * mm
B_a = 0.1 * tesla

def wire_field(x, y, z):
    rho2 = x**2 + y**2
    if rho2 == 0.0:
        return [0.0, 0.0, 0.0]
    return [-B_a * a * y / rho2, B_a * a * x / rho2, 0.0]

field = fields.CustomMagneticField(name="wire")
field.field_function = wire_field
box.add_field(field)
\end{lstlisting}

\subsection{Validation tests}
\label{sec:validation}

We validate the framework with three sets of tests targeting its compatibility with Geant4, its handling of coordinate transformations and dynamic geometry, and the equivalence of the three different field providers. Additionally, we quantify the computational cost of each provider. All tests track \SI{100}{\mega\electronvolt} protons through the fields in vacuum. Vacuum is chosen because in the absence of interactions with matter, the trajectories are determined by the field alone, so that any difference between two simulations can be attributed to the field implementation rather than to stochastic interactions.

Unless otherwise stated, all tests use the same baseline geometry, phase-space input, physics list and tracking setup. The geometry consists of a \SI{4}{\metre} cubic vacuum world containing a \SI{500}{\milli\metre} vacuum cube centred at the origin, which serves as the host volume for the field. The beam consists of \num{10000} protons with \SI{100}{\mega\electronvolt} kinetic energy, emitted from a disk of \SI{50}{\milli\metre} radius with positions distributed uniformly across the disk and directions distributed uniformly within a cone of \SI{10}{\degree} half-angle. In vacuum, each trajectory is fully determined by the initial position and direction of the proton and by the field, so no statistical averaging is needed. The protons start from different positions and directions so that their trajectories sample a large part of the field region, giving between \num{1.7e6} and \num{2.1e6} compared step points per pair of simulations. The phase-space source is located \SI{250}{\milli\metre} upstream of the host volume. All simulations use the \texttt{QGSP\_BERT\_EMV} reference physics list. Unless otherwise specified, transport steps inside the host volume are limited to \SI{3}{\milli\metre}, and all simulations are run single-threaded with the same random seed.

All simulations were run on compute nodes of the Merlin7 cluster at the Paul Scherrer Institute, each equipped with two 64-core AMD EPYC~7742 processors and running SUSE Linux Enterprise Server~15~SP6. They used the GATE~10 development version at commit \texttt{0bc96f8} (following release~10.1.1) and Geant4~11.4.2 built with multithreading support, both compiled in release mode with GCC~7.5.0, together with CPython~3.14.3 in its standard build. The scripts of all tests are available in \citet{ballestero2026validation}.

For trajectory-equivalence tests, the trajectories of the primary protons are compared at each recorded step point using the differences in particle position, direction, and kinetic energy. For each quantity, the maximum difference across all step points and primaries is reported as the maximum residual. Trajectories are considered \emph{bitwise identical} when every recorded double-precision quantity agrees exactly, i.e. all residuals are zero.

\paragraph{Equivalence with native Geant4.}
We first test whether the framework preserves the native Geant4 tracking. A bare Geant4 application is compared with a GATE~10 simulation using the framework for the four field types with built-in Geant4 counterparts: uniform magnetic, ideal quadrupole, ideal sextupole, and uniform electric fields. The host volume is placed at the world origin without rotation, reducing the coordinate transformation pipeline to the identity and isolating the field evaluation interface.

The two applications use the same physics list and tracking configuration, including the parameters of Table~\ref{tab:fieldparams}. Both applications are built against the same Geant4 installation with the same compiler and optimisation flags, and are run single-threaded on the same machine.

The four fields are $\mathbf{B}_\mathrm{uniform} = (0.1,0.5,-0.2)\,\si{\tesla}$, $\mathbf{B}_\mathrm{quadrupole} = g(y,x,0)$ with $g=\SI{5}{\tesla\per\metre}$, $\mathbf{B}_\mathrm{sextupole} = g(xy,(x^2 -y^2)/2,0)$ with $g=\SI{5}{\tesla\per\metre\squared}$, and $\mathbf{E}_\mathrm{uniform} = (0.1,0.5,-0.2)\,\si{\mega\volt\per\metre}$.

\paragraph{Coordinate handling and dynamic geometry.}
We next test whether a field defined in the local coordinate system of a host volume is preserved under geometry transformations and dynamic motion. All tests use an ideal magnetic quadrupole with gradient $g=\SI{5}{\tesla\per\metre}$ defined in the local coordinate system of the host volume.

First, the host volume is placed at the world origin in one configuration, and then it is translated by $(60,-40,0)\,\si{\milli\metre}$ and rotated by $(30,40,50)\,\si{\degree}$ around the $x$, $y$, and $z$ axes in a second configuration. The particle source undergoes the same rigid transformation, so the two configurations represent the same physical situation in different world frames. The resulting trajectories should therefore be related by the applied rigid transformation. The transformed trajectory is brought back to the reference frame before calculating the residual.

Second, a spherical daughter volume of radius \SI{100}{\milli\metre} is placed at the centre of the host volume. The daughter is rotated by $(20,-35,45)\,\si{\degree}$ in one configuration and left unrotated in the other. Because the spherical boundary is invariant under rotation and the daughter carries no field of its own, the field should be unchanged. This test therefore isolates the effect of the additional navigation level between the particle and the field host introduced by the daughter volume.

Finally, the host volume is translated and rotated over three consecutive runs using a dynamic actor. The resulting trajectories are compared with those obtained from three static simulations representing the same placements, while the beam remains fixed. This test verifies that the field follows the physical placement of its host volume when the geometry is modified between runs.

\paragraph{Provider equivalence and cost.}
We finally evaluate the equivalence and computational cost of the three field providers. For the equivalence test, an ideal magnetic sextupole $\mathbf{B}=g\left(xy,\;(x^2-y^2)/2,\;0\right)$ with $g=\SI{1e-4}{\tesla\per\square\milli\metre}$ is defined in the local frame of the host volume. A sextupole was selected because its quadratic dependence on position cannot be reproduced exactly by trilinear interpolation, unlike uniform and quadrupole fields. The analytic provider uses the compiled closed form, while the callback provider evaluates a Python function returning the same expression. The mapped provider samples the field on Cartesian grids of $126^3$, $251^3$ and $501^3$ nodes spanning the host volume, corresponding to node spacings of \SIlist{4;2;1}{\milli\metre}, with both nearest-neighbour and trilinear interpolation tested at every spacing. Trajectory residuals are evaluated against the analytic provider, which is taken as the reference.

For the runtime measurements, the phase-space source is generated with \num{1000000} primaries using the same energy, spatial, and angular distributions as the baseline source. The field is the uniform magnetic field $\mathbf{B}=(0.1,0.5,-0.2)\,\si{\tesla}$, which the mapped provider samples on a grid of $100^3$ nodes spanning the host volume. The wall-clock time of the event loop, which excludes the initialisation of geometry, physics, fields, and worker threads, is measured using one, two, four, and eight worker threads, with five independent runs for each configuration. The median runtime and the full range of the five runs are recorded for each provider. The number of primaries is chosen so that the tracking time remains well above the fixed per-run overhead still included in the measurement.

Each run of one provider at all four thread counts was executed as a single job with exclusive access to a compute node, so that no other workload shared the node during the measurement. The five runs of each provider were executed on different nodes, so the reported ranges include node-to-node variation.
Each job first executed a discarded single-threaded simulation of \num{200} primaries to warm the file-system caches.

\section{Results}
\label{sec:results}

\begin{table}[htbp]
\centering
\caption{Maximum residuals of the trajectory-equivalence tests, taken
over every recorded step point of every primary. For the provider-equivalence
tests, the analytic provider serves as the reference and the field is an
ideal sextupole of gradient \SI{1e-4}{\tesla\per\square\milli\metre}, which
neither interpolation scheme reproduces exactly; the mapped provider is
therefore listed once per grid, the node spacings $h$ of
\SIlist{4;2;1}{\milli\metre} corresponding to \num{126}, \num{251} and
\num{501} nodes per axis across the host volume. These tests use the baseline geometry, source and
tracking setup. A residual of exactly \num{0} denotes bitwise identity of every
recorded quantity.}
\label{tab:residuals}
{\small
\begin{tabular}{
    @{}
    >{\raggedright\arraybackslash}p{3.2cm}
    l
    S[table-format=1.2e-2]
    S[table-format=1.2e-2]
    S[table-format=1.2e-2]
    @{}
}
\toprule
Test & Configuration
  & {$\max \Delta r$ [\si{\milli\metre}]}
  & {$\max \Delta \theta$ [\si{\radian}]}
  & {$\max \Delta E_k$ [\si{\mega\electronvolt}]} \\
\midrule
Native Geant4 & Uniform $\mathbf{B}$ & 0 & 0 & 0 \\
 & Quadrupole $\mathbf{B}$ & 0 & 0 & 0 \\
 & Sextupole $\mathbf{B}$ & 0 & 0 & 0 \\
 & Uniform $\mathbf{E}$ & 0 & 0 & 0 \\
\addlinespace
\addlinespace
Coordinate transforms & Translation + rotation & 1.07e-12 & 3.01e-15 & 0 \\
 & Nested volume & 1.64e-12 & 1.00e-15 & 0 \\
 & Dynamic geometry & 0 & 0 & 0 \\
\addlinespace
\addlinespace
Provider equivalence & Mapped, nearest, $h=\SI{4}{\milli\metre}$ & 1.20e+00 & 5.87e-03 & 0 \\
 & Mapped, nearest, $h=\SI{2}{\milli\metre}$ & 3.99e-01 & 1.78e-03 & 0 \\
 & Mapped, nearest, $h=\SI{1}{\milli\metre}$ & 1.69e-01 & 7.88e-04 & 0 \\
 & Mapped, trilinear, $h=\SI{4}{\milli\metre}$ & 1.15e-02 & 4.94e-05 & 0 \\
 & Mapped, trilinear, $h=\SI{2}{\milli\metre}$ & 2.66e-03 & 1.15e-05 & 0 \\
 & Mapped, trilinear, $h=\SI{1}{\milli\metre}$ & 5.58e-04 & 2.39e-06 & 0 \\
 & Callback & 0 & 0 & 0 \\
\bottomrule
\end{tabular}}
\end{table}

\subsection{Compatibility with native Geant4}

The residuals of every trajectory-equivalence test are collected in Table~\ref{tab:residuals}. For the four field types with native Geant4 counterparts, the GATE~10 trajectories were bitwise identical to the corresponding bare Geant4 ones, with exactly zero residuals in position, direction, and kinetic energy for the uniform magnetic field, quadrupole, sextupole, and uniform electric field.

\subsection{Coordinate transformations and dynamic geometry}

The combined translation and rotation test yielded a maximum position residual of \SI{1.07e-12}{\milli\metre} and a maximum directional residual of \SI{3.01e-15}{\radian}, while the nested-volume test yielded \SI{1.64e-12}{\milli\metre} and \SI{1.00e-15}{\radian}, respectively. The trajectories of the dynamic-geometry test were bitwise identical to the corresponding static configurations. No difference in kinetic energy was observed in any of the tests. The non-zero residuals were at the floating-point precision level and did not correspond to a physically relevant difference in the tested trajectories.

\subsection{Provider equivalence and cost}

The three field providers also reproduced equivalent transport for the common sextupole field. The callback provider, which evaluates the same closed form in Python, produced trajectories bitwise identical to those of the analytic provider in position, direction and kinetic energy. The mapped provider converged towards the analytic reference as the grid was refined: the maximum position residual of trilinear interpolation fell from \SI{1.15e-2}{\milli\metre} at a node spacing of \SI{4}{\milli\metre} to \SI{5.58e-4}{\milli\metre} at \SI{1}{\milli\metre}, and that of nearest-neighbour interpolation from \SI{1.20}{\milli\metre} to \SI{1.69e-1}{\milli\metre}. At equal resolution, trilinear interpolation was therefore between \num{105} and \num{304} times more accurate than nearest-neighbour interpolation. No difference in kinetic energy was observed in any configuration, as expected for a purely magnetic field.

\begin{figure}[tbp]
 \centering
 \includegraphics{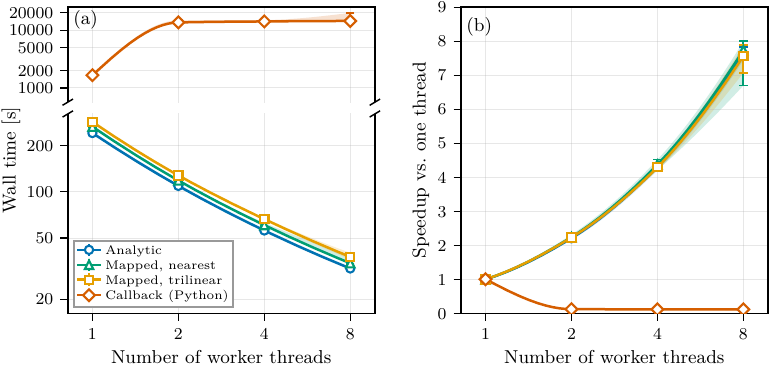}
 \caption{Cost of the three field providers for the tracking of \num{1000000} primaries, shown as the median of five runs, with bands and error bars showing the full range of the runs. (\emph{a}) Wall-clock time of the tracking phase against the number of worker threads. (\emph{b}) The same data as speedup with respect to the single-threaded run of the same provider.}
 \label{fig:cost}
\end{figure}

The cost of the three providers is shown in Figure~\ref{fig:cost}. Single-threaded, the analytic provider tracked \num{1000000} primaries in \SI{241.4}{\second} (range \SIrange{240.1}{247.8}{\second}). Relative to it, the mapped provider required \SI{9.9}{\percent} more runtime with nearest-neighbour interpolation and \SI{18.1}{\percent} more with trilinear interpolation, while the callback provider required \num{6.87} times as much. The analytic provider remained the fastest at every thread count.

The providers differed also in their multi-threaded behaviour. The compiled providers continued to scale over the whole range tested, reaching speedups of \numrange{2.21}{2.24} at two worker threads, \numrange{4.30}{4.37} at four and \numrange{7.56}{7.74} at eight. The callback provider instead became slower as threads were added, requiring \num{8.2} times its single-threaded runtime at two threads and \num{8.7} times at eight threads. Its runtime relative to the analytic provider consequently increased from \num{6.9} to \num{455}.

\section{Discussion}
\label{sec:discussion}

The proposed framework integrates EM fields into the GATE geometry interface without modifying the underlying Geant4 particle transport. The framework intervenes only at the field-evaluation interface exposed by Geant4, while the equation of motion, numerical integration and boundary-intersection machinery remain native Geant4 components. The bitwise agreement obtained for field types with native Geant4 counterparts demonstrates that this additional GATE layer does not alter the underlying transport. At the same time, separating coordinate handling and field evaluation from the transport machinery allows field representations to be extended independently. This provides a stable integration point with Geant4, so that changes to the underlying tracking algorithms do not require a separate reimplementation on the GATE side.

Another key aspect is that fields can be defined relative to the physical component to which they belong rather than in the global simulation frame. This is particularly relevant for geometries containing EM elements with their own position and orientation, since the same field definition can then be reused for different physical placements. The validation results show that this association is preserved through translations, rotations and additional levels of geometry nesting, and that it remains valid when the geometry is modified between simulation runs. This naturally couples the field definition to the geometry and avoids requiring users to manually transform field definitions when EM components are repositioned or rotated. The same mechanism also allows a field attached to a logical volume to be reused by multiple physical placements, while retaining the local orientation of each placement.

The three-provider architecture provides users with flexibility in how EM fields are defined while allowing the computational cost of field evaluation to be tailored to the intended application. Analytic fields provide the lowest evaluation cost when a suitable closed form is available. Mapped fields extend this approach to arbitrary spatial distributions supplied on regular grids: in the validation tests, the additional cost was modest when compared with the analytic provider, including when trilinear interpolation was used. They therefore provide a practical option for production simulations in which fields are obtained from numerical solvers or measurements. Callback fields offer the greatest flexibility, allowing arbitrary spatial dependencies to be expressed directly in Python. This flexibility, however, comes at a substantially higher computational cost, which scales poorly with the number of worker threads, a behaviour consistent with field evaluation being serialised by the global interpreter lock and with contention for that lock. Callback fields are therefore best suited for prototyping and exploratory work, while analytic and mapped fields are more appropriate for computationally intensive simulations.

The validation tests were designed to test the principal architectural properties of the framework rather than to exhaustively cover all possible field definitions and geometries. The trajectory tests included representative native Geant4 field types, rigid transformations, nested volumes and dynamic geometry, while the provider tests demonstrated equivalent transport for a common field and quantified the computational cost of the different implementations. All tests were performed in vacuum, so that the field was the only influence on the trajectories. This choice does not limit the results to transport in vacuum: the framework only supplies the field value requested by the Geant4 integrator, while energy loss, scattering and the production of secondary particles in matter remain handled by the unmodified Geant4 physics models, as in any Geant4 simulation with a field. The performance measurements were obtained for a controlled workload and should therefore not be interpreted as universal timing benchmarks: the relative cost of field evaluation will depend on field complexity, the number of field queries, the tracking configuration and the materials traversed. These limitations should be considered when selecting a provider for a particular application.

\section{Conclusion}
\label{sec:conclusion}

We introduced a framework for defining EM fields in GATE~10 and associating them directly with simulation volumes while retaining the native Geant4 particle-transport machinery. The framework separates geometry-dependent coordinate handling from field evaluation and provides a common interface for analytic, mapped, and Python callback fields. The validation tests demonstrated compatibility with native Geant4 tracking, correct behaviour under translations, rotations, nested volumes, and dynamic geometry, and equivalent transport across the available field providers. Analytic and mapped fields provide efficient options for production simulations, while callback fields offer greater flexibility for prototyping and exploratory applications. Our contribution has been included in the GATE~10 codebase, and is therefore open source and publicly available to researchers. The developed architecture also allows additional field representations to be introduced without modifying the transport or coordinate-handling layers, providing an extensible basis for EM field transport in GATE~10. By coupling fields to the simulation geometry, the framework is well suited to medical-physics applications that involve charged-particle transport in EM fields, for example simulations of MRgRT, of treatment beamlines and rotating gantries containing several independently positioned magnets, or of detectors with fields confined to individual components, such as Faraday cups for proton dosimetry.

\ack{The authors would like to thank Clémence Gentner for her collaboration in the earliest prototypes of this framework. We also thank the GATE community and coordinators for their support and feedback during the development process, as well as for their open and inclusive approach to research software development.}

\funding{This research was funded in whole or in part by the Swiss National Science Foundation (SNSF) [grant number \href{https://data.snf.ch/grants/grant/10002635}{10002635}]. For the purpose of Open Access, a CC BY public copyright licence is applied to any Author Accepted Manuscript (AAM) version arising from this submission.}

\roles{

\begin{itemize}
\item \textbf{Marc Ballestero Ribó:} Conceptualisation, Methodology, Software, Validation, Formal analysis, Investigation, Visualisation, Writing -- original draft.
\item \textbf{Carla Winterhalter:} Conceptualisation, Validation, Writing -- review \& editing.
\item \textbf{Antony John Lomax:} Conceptualisation, Resources, Writing -- review \& editing, Project administration.
\item \textbf{Sairos Safai:} Conceptualisation, Resources, Writing -- review \& editing,  Project administration, Funding acquisition.

\end{itemize}
}
\data{The framework described in this article is part of GATE~10, which is open source and publicly available at the \href{https://github.com/OpenGATE/opengate}{OpenGATE GitHub repository}. It was contributed through pull requests \href{https://github.com/OpenGATE/opengate/pull/1003}{\#1003}, \href{https://github.com/OpenGATE/opengate/pull/1012}{\#1012}, \href{https://github.com/OpenGATE/opengate/pull/1112}{\#1112} and \href{https://github.com/OpenGATE/opengate/pull/1130}{\#1130}, all included in commit \texttt{0bc96f8}, which was used for this work. Documentation and examples are available in the \href{https://opengate-python.readthedocs.io/en/master/user_guide/user_guide_fields.html}{GATE~10 user guide}. The code of the validation tests and the results reported in this article, including the data needed to reproduce Table~\ref{tab:residuals} and Figure~\ref{fig:cost}, will be made publicly available at \url{https://doi.org/10.5281/zenodo.23685931} \citep{ballestero2026validation}.}

\section*{References}
\printbibliography[heading=none]

\end{document}